\documentclass[a4paper]{revtex4}

\usepackage{bm}
\usepackage{amsmath}
\usepackage{amssymb}
\usepackage{amsfonts}
\newcommand\Def{
  \,{\buildrel {\rm def} \over =}\,
}
\newcommand{\pos}{q}
\newcommand{\kin}{\pi}
\renewcommand{\ss}{{0}}
\newcommand{\dm}[1]{\d\{#1\}} 
\newcommand{\sym}[1]{\left[#1\right]_{\rm sym}}
\newcommand{\stsp}{\Sigma}
\newcommand{\ti}{\theta}
\newcommand{\op}{\hat}
\newcommand{\atiop}{^\vee}
\newcommand{\tiop}{\check}
\newcommand{\D}{{\mathcal D}}
\renewcommand{\H}{{\mathcal H}}
\newcommand{\Prob}{{\rm Prob}}
\newcommand{\be}[1]{\begin{equation}\label{#1}}
\newcommand{\ee}{\end{equation}}
\newcommand{\ba}[1]{\begin{array}{#1}}
\newcommand{\ea}{\end{array}}
\renewcommand{\d}{{\rm d}}
\newcommand{\lv}{\langle}
\newcommand{\rv}{\rangle}

\newcommand{\lp}{\left(}
\newcommand{\rp}{\right)}
\renewcommand{\O}{{\mathcal O}}

\begin{document}
\title{On the Gallavotti-Cohen symmetry for stochastic systems}
\author{Martin Depken}
\affiliation{University of Oxford, Department of Physics, Theoretical Physics\\ 1 Keble Road, Oxford, OX1 3NP, U.K.}
\begin{abstract}
Considering Langevin dynamics we derive the general form of the stochastic differential that satisfies the Gallavotti-Cohen symmetry. This extends the work previously done by Kurchan, and Lebowitz and Spohn on such systems, and we treat systems with and without inertia in a unified manner.  We further shown that for systems with a time-reversal invariant steady state there exists a stochastic differential for which then the Gallavotti-Cohen symmetry, and all its consequences, are valid for finite times. For these systems the differential can be seen as the direct analogy of the Gibbs-entropy creation along paths in deterministic systems. It differs from previously studied differentials in that it identically zero for equilibrium systems while on average strictly positive for non-equilibrium system. When the steady state is not time reversal invariant the Gallavotti-Cohen symmetry is asymptotically valid in the usual long time limit.
\end{abstract}
\maketitle
\section{Introduction}
In closed and isolated  Hamiltonian systems the Liouville theorem states that the volume of any phase-space element is invariant under the time evolution, i.e. the phase-space contraction rate, $\sigma(\bm\pos^t,\bm\kin^t)$, vanishes. If the system couples to the outside world this is in general no longer true. The phase-space contraction rate fluctuates along trajectories, and it has been shown~\cite{Andrej82,GallavottiBOOK} that this rate can be interpreted as the Gibbs-entropy creation rate. In~\cite{Gallavotti95a} it was further shown that if a system satisfies the chaotic hypothesis~\cite{Gallavotti96,GallavottiBOOK}, then the probability distribution for the average phase-space contraction along a realisation, $\bar \sigma$,  satisfies
$$
  \frac{\Prob(\bar\sigma)}{\Prob(-\bar\sigma)}\sim \exp(T\bar\sigma), \quad \bar \sigma=\frac{1}{T}\int_{0}^T\!\d t\,\sigma({\bm\pos}^{t},{\bm\kin}^{t}), \quad T\rightarrow \infty.
$$
This relationship is called the fluctuation theorem and was initially observed in numerical simulations~\cite{Evans93}. This led to the formulation of the  chaotic hypothesis, which enabled a theoretical derivation of the above result (see \cite{GallavottiBOOK} and references therein). This theorem is especially interesting since it is a parameter free prediction, valid arbitrarily far from equilibrium, concerning quantities with a potentially physical interpretation. The appreciation of its importance was further increased when it was realised~\cite{Gallavotti96} that it gives the fluctuation-dissipation theorem of equilibrium statistical mechanics in the linear regime around equilibrium. 

The fluctuation theorem can also be written as a symmetry relation of the moment-generating functional of the phase-space contraction along paths, and we follow~\cite{Lebowitz99} in referring to this as the Gallavotti-Cohen symmetry.

We here consider stochastic systems evolving with Langevin dynamics. Even though stochastic systems have already been considered in this setting~\cite{Kurchan98,Lebowitz99,Maes03}, we here slightly generalize the explicit derivations, and treat the systems under consideration in a unified manner. We further show that under certain circumstances we can define a stochastic differential that satisfies the fluctuation theorem for finite times. This stochastic differential is also shown to be on average strictly positive for non-equilibrium systems, while it vanishes identically for equilibrium systems. By analogy with the above mentioned situation for deterministic systems we therefore suggest identifying this stochastic differential with the Gibbs-entropy creation along a infinitesimal step of the evolution.

In what follows we will explicitly examine the effect of time reversal on the level of path probability densities, and use this to derive the Gallavotti-Cohen symmetry. This puts the derivation for system with and without inertia on an equal footing.

\section{The Langevin process}
\label{sec:lang}
We consider a collection of fields, $\phi=\{\phi_\alpha\}\in\stsp$, evolving through the state space $\stsp$ with stochastic dynamics given by the Langevin equation
\be{eq:lang}
 \d_{\rm s}\phi_\alpha=f_\alpha(\phi)\d t+ \Gamma_{\alpha i}(\phi)\d_{\rm s} w_i.
\ee
The differential $\d_{\rm s}$ is interpreted as a Stratonovich differential, which corresponds to the limit of a finite noise-correlation time approaching zero~\cite{Stratonovich66}. The index $\alpha$ is taken to be discrete, continuous, or mixed, and we sum/integrate over repeated indexes. The noise term $\d_{\rm s} w_i(t)$ is constructed from a standard, uncorrelated, Wiener process. Since the noise originates in another physical space (the heat reservoir) we choose to use Latin subscripts to enumerate its components. We further assume the noise-correlation matrix
$$
  \Xi_{\alpha\beta}(\phi)=\frac{1}{2}\Gamma_{\alpha i}(\phi)\Gamma_{\beta i}(\phi) 
$$
to be positive definite for all $\phi\in\stsp$. For later reference we note that the Stratonovich interpretation gives the normal differential rule
$$
 \d_{\rm s}g(\phi)=\partial_\alpha g(\phi)\d_{\rm s}\phi_\alpha,\quad  \partial_\alpha\Def\frac{\partial}{\partial \phi_\alpha}.
$$
\subsection{Time reversal}
\label{sec:trev}
Since one often models Hamiltonian systems connected to heat reservoirs with an equation like~(\ref{eq:lang}), we need to allow for degrees of freedom that transform under the reversal of the direction of time. Given that a Hamiltonian is an even function in the generalized momentums, Hamilton's equations are invariant under the time-reversal transformation
$$
  t\rightarrow -t, \quad \bm\pos \rightarrow \bm\pos, \quad \bm\kin\rightarrow -\bm\kin.
$$
When coupling the Hamiltonian system to the external world, or when we consider coarse-grained degrees of freedom, we will no longer assume the dynamics to be time-reversal invariant. This because we have given up on a complete description of the system, and any time reversal that can be observed will only effect our coarse-grained degrees of freedom ($\phi$), and the driving. That is, we assume the stochastic effects of the noise to be invariant under time reversal. We now introduce a generalized time-reversal map $\ti$ such that
$$
  \ti\phi=\{\ti_{\alpha}(\phi)\}, \quad  \ti^2=1 \quad \Rightarrow \quad\d_{\rm s}(\ti\phi_\alpha)=\ti_{\alpha\beta}(\phi)\d_{\rm s}\phi_{\beta}, \quad \ti_{\alpha\gamma}(\ti\phi)\ti_{\gamma\beta}(\phi)=\delta_{\alpha\beta},
$$
where $\ti_{\alpha\beta}(\phi)\Def\partial_{\beta}\ti_\alpha(\phi)$. In what follows we will omit the field dependence of $\ti_{\alpha\beta}$ since it always corresponds to the filed dependence of the vector it is applied to. The effect of time reversal on the evolution equation is
$$
  \left. \ba{r}\phi\rightarrow\ti\phi\\ t\rightarrow - t \ea\right\}\quad \Rightarrow \quad \left\{\ba{c}\d_{\rm s}\phi_\alpha=f_\alpha(\phi)\d t+\Gamma_{\alpha i}(\phi)\d_{\rm s}w_i\\ \rightarrow\\ \ti_{\alpha\beta}(\phi)\d_{\rm s}\phi_\beta=f_{\beta}(\ti \phi)(-\d t)+\Gamma_{\beta i}(\ti \phi)(-\d_{\rm s} w_i).\ea\right.
$$
Since the stochastic effect of the noise is taken to be time-reversal invariant, we require $\Gamma_{\alpha i}(\ti\phi)=\ti_{\alpha\beta}\Gamma_{\beta i}(\phi)$. It is convenient to introduce the reversible and irreversible parts of any vector $h_\alpha(\phi)$ according to
$$
  h_\alpha^{\rm R}(\phi)\Def \frac{1}{2}( h_\alpha(\phi)-\ti_{\alpha\beta} h_\beta(\ti \phi)), \quad h_\alpha^{\rm I}(\phi)\Def\frac{1}{2}(h_\alpha(\phi)+\ti_{\alpha\beta} h_\beta(\ti\phi)),
$$
which transform as
$$
  h_\alpha^{\rm R}(\ti\phi)=- \ti_{\alpha\beta}h_\beta^{\rm R}(\phi), \quad  h_\alpha^{\rm I}(\ti\phi)= \ti_{\alpha\beta} h_\beta^{\rm I}(\phi).
$$
With these definition the evolution equations take the form
$$
\ba{lrl}
\textrm{Original:}&\d_{\rm s}\phi_\alpha&=(f^{\rm R}_{\alpha}( \phi)+f^{\rm I}_{\alpha}( \phi))\d t+\Gamma_{\alpha i}(\phi)\d_{\rm s} w_i,\\
\textrm{Time reversed:~~}&\d_{\rm s} \phi_\alpha&=(f^{\rm R}_{\alpha}( \phi)-f^{\rm I}_{\alpha}( \phi))\d t-\Gamma_{\alpha i}(\phi)\d_{\rm s} w_i,
\ea
$$
and we see that the equations of motion are time-reversal invariant only if $f_\alpha^{\rm I},\Gamma_{\alpha i}\equiv 0$.
\subsection{The Fokker-Planck equation and an operator formalism}
Under the dynamics given by~(\ref{eq:lang}) a probability density, $\rho$, will evolve according to the Fokker-Planck equation,
\be{eq:fokker}
  \partial_t \rho(\phi,t)=-\partial_\alpha J_\alpha(\phi,t), \quad J_\alpha(\phi,t)= k_\alpha(\phi)\rho(\phi,t)-\partial_\beta(\Xi_{\alpha\beta}(\phi)\rho(\phi,t)),
\ee
where $k_\alpha(\phi)$ is a drift vector, given by
\be{eq:drift}
k_\alpha(\phi)= f_\alpha(\phi)+\frac{1}{2} \Gamma_{\beta i}(\phi)\partial_\beta \Gamma_{\alpha i}(\phi).
\ee 
Viewing~(\ref{eq:fokker}) as a conservation equation for probability, we interpret $J_\alpha(\phi,t)$ as the probability current through state space. Introducing the generalized potential 
$$
  \Phi(\phi,t)\Def-\ln\rho(\phi,t),
$$
we can also define the streaming velocity in phase space, $j_\alpha(\phi,t)$, through
\be{eq:altfokk}
  J_{\alpha}(\phi,t)= j_\alpha(\phi,t)\rho(\phi,t), \quad j_\alpha(\phi,t)= k_\alpha(\phi)+\Xi_{\alpha\beta}(\phi)\partial_\beta \Phi(\phi,t)-\partial_\beta\Xi_{\alpha\beta}(\phi).
\ee
From~(\ref{eq:fokker}) and (\ref{eq:altfokk}) it follows that
$$
  \partial_t \lv \phi_\alpha(t)\rv=-\lv j_\alpha(\phi(t),t)\rv=-\lv k_\alpha(\phi(t)) \rv.
$$
Looking at~(\ref{eq:drift}), it is clear that the evolution of the averaged fields contain contributions from both deterministic and stochastic influences (c.f. the behaviour of a driven Brownian particle in an non-uniform temperature field).

To facilitate the transition to a path-integral formalism we put the Fokker-Planck equation in a bra-ket notation.  Assume that we have a Hilbert space, spanned by the orthonormal basis $\{|\phi\rv\}_{\phi\in\stsp}$, and having the usual quadratic norm. In this space we represent probability densities with
$$
  |\rho(t)\rv\Def\int_\stsp\dm \phi |\phi\rv \rho(\phi,t), \quad \dm \phi\Def\prod\limits_\alpha\d \phi_\alpha,
$$
and define the field operators $\op \phi_\alpha$ and their canonical conjugates $\op p_\alpha$ through
$$
 \op \phi_\alpha\Def\int_\stsp \dm \phi |\phi\rv \phi_\alpha \lv \phi|,\quad\op p_\alpha\Def\int_\stsp \dm \phi |\phi\rv \imath \partial_\alpha \lv \phi|.
$$
These operators satisfy the usual canonical commutator relation \mbox{$[\op p_\alpha,\op \phi_\beta]=\imath\delta_{\alpha\beta}$}. For the time reversed field operators we introduce the notation
$$
  \tiop \phi_\alpha=\ti_{\alpha}\op \phi\quad \textrm{and}\quad \tiop p_\alpha=\ti_{\beta\alpha}\op p_\beta.
$$
From the above it directly follows that for a differentiable function, $g(\phi)$, we  have
\be{eq:commu}
  [\op p_\alpha,\op g]=\imath\partial_{\alpha}\op g, \quad  [\tiop p_\alpha,\tiop g]=\imath(\partial_{\alpha}g)\atiop=\imath\ti_{\beta\alpha}\partial_{\beta}\tiop g,
\ee
where we have made use of the short hand notation $\op g=g(\op \phi)$ and $\tiop g=g(\tiop \phi)$. In this operator notation the Fokker-Planck equation can conveniently be written as
$$
  \partial_t|\rho(t)\rv = \op \Omega |\rho(t)\rv, \quad   \op \Omega =\imath \op p_\alpha \op k_\alpha  - \op p_\alpha\op p_\beta \op\Xi_{\alpha\beta}.
$$
We also define the left vector
$$ 
  \lv 0|\Def\int_\stsp \d{\phi}\lv \phi|,
$$
which satisfies $\lv 0|\op\Omega=0$ for all probability conserving evolution operators $\op\Omega$. Using the operator notation we can write conditional probabilities as
\be{eq:oprep}
  \Prob(\phi',t'|\phi,t)=\lv \phi'|\exp((t'-t)\op \Omega)|\phi\rv.
\ee
We are interested in the Stratonovich interpretation of stochastic differentials, and it turns out to be advantageous to write operators in a symmetric fashion with respect to $\op \phi$ and $\op p$. More precisely, we put any operator $\op A=A(\op \phi,\op p)$, which is a finite polynomial in $\op p$, in the form
$$
  \op A= \sum_{n=1}^{\cdot} \frac{1}{2}\lp\op p_{\alpha_1}\cdots\op p_{\alpha_n} a^n_{\alpha_1,\ldots,\alpha_n}(\op \phi)+a^n_{\alpha_1,\ldots,\alpha_n}(\op \phi)\op p_{\alpha_1}\cdots\op p_{\alpha_n}\rp+a^0(\op \phi).
$$
Here the quantities $a^n_{\alpha_1,\ldots,\alpha_n}(\op \phi)$ are all independent of $\op p$, and can be calculated through the use of the commutator relationship~(\ref{eq:commu}). To simplify notation we use the above to introduce the ordering operation $\sym{\cdot}$ through
$$
  \sym{\sum_{n=1}^{\cdot}\op p_{\alpha_1}\cdots\op p_{\alpha_n} a^n_{\alpha_1,\ldots,\alpha_n}(\op \phi)+a_0(\op \phi)}\Def \op A,
$$
with the help of which the evolution operator can be written as  
\be{eq:symomega}
  \op\Omega=\Omega_{\rm sym}(\op \phi,\op p)\Def\sym{\imath \op p_\alpha (\op k_\alpha-\partial_\beta \op\Xi_{\alpha\beta})-\op p_\alpha\op p_\beta \op\Xi_{\alpha\beta} -\frac{1}{2}\partial_\alpha\op k_\alpha}.
\ee
\subsection{Detailed balance}
\label{sec:db}
For a closed and isolated Hamiltonian system it can be shown that the transition probabilities and the steady-state (equilibrium) probability density, $\rho_{\ss}$, satisfy the detailed-balance condition~\cite{Kampen81b,Graham71}
\be{eq:db1}
  \Prob(\phi',t'|\phi,t)\rho_{\ss}(\phi)= \Prob(\ti\phi,t'|\ti\phi',t)\rho_{\ss}(\ti\phi').
\ee
On the right hand side of the above we have implicitly time reversed all external quantities as well (such as external magnetic fields). The relationship~(\ref{eq:db1}) is a direct consequence of the time-reversal invariance exhibited by the equations of motion.  Considering every order of $t-t'$ in the operator representation~(\ref{eq:oprep}), (\ref{eq:db1}) is equivalent to
\be{eq:dbop}
  \op\rho_{\ss}= \tiop\rho_{\ss}, \quad \op \Omega \op \rho_{\ss}=\tiop\rho_{\ss}\tiop\Omega^\dagger.
\ee
The evolution equation we are considering is not in general time-reversal invariant, and thus such a system will in general not satisfy detailed balance. It was shown in~\cite{Kampen81b,Graham71} that for a system like the one considered here, the detailed-balance condition is equivalent to the potential conditions
\be{eq:pot}
\ba{c}
   \Phi_\ss(\phi)=\Phi_\ss(\ti \phi) ,\quad J_\alpha^{\rm I}(\phi)=0,\quad \partial_\alpha  J_\alpha^{\rm R}(\phi)=0\\ \Leftrightarrow\\\Phi_\ss(\phi)=\Phi_\ss(\ti \phi) ,\quad j_\alpha^{\rm I}(\phi)=0, \quad(\partial_\alpha-\partial_\alpha\Phi_0(\phi))j_{\alpha}^{\rm R}(\phi)=0,
\ea
\ee
where we have used the generalized potential $\Phi_\ss=-\ln\rho_\ss$. The condition on the reversible current is always satisfied in a steady state, and thus the detailed-balance condition is equivalent to requiring the absence of any irreversible currents, and that the steady-state probability density is time-reversal invariant.

\section{The Gallavotti-Cohen symmetry through path integrals}
In the proof of the fluctuation theorem for thermostatted Hamiltonian systems~\cite{Gallavotti95a}, time reversal played a crucial part in that it enabled the  authors to write the phase-space contraction along a path in terms of the phase-space contraction along the corresponding time-reversed path. Following this in spirit, we will consider the effect of time reversal explicitly on the path probability densities, and see how it gives the Gallavotti-Cohen symmetry also for stochastic systems.
\subsection{Path integrals}
\label{sec:pathint}
A path-integral formulation for the Langevin process was pioneered by Onsager and Machallup \cite{Onsager53,Machlup53} when considering a linear Langevin equation. We will derive a similar formulation for the fully non-linear case, but only push the derivations as far as will be needed in order to derive the Gallavotti-Cohen symmetry (for a full treatment see~\cite{Graham77}). 

The system is assumed to be in a steady state with probability density $\rho_\ss$. Using the operator formalism introduced in the previous chapter we can move to a standard path-integral formalism, much in the same way as is done in quantum mechanics. We slice the time direction in intervals of size $\delta t$ and let $t_k=\delta t k$. With the help of~(\ref{eq:oprep}), we write the probability density of a discrete sampling, $\phi_{\delta t}(\cdot)=(\phi(t_1),\phi(t_2),\ldots,\phi(t_{T/\delta t}))$, of the realization as
$$
  \Prob(\phi_{\delta t}(\cdot))=\left[\prod_{k=0}^{T/\delta t-1}\lv \phi(t_{k+1})|\exp(\delta t\, \op\Omega)|\phi(t_k)\rv\right] \exp(-\Phi_\ss(\phi(0))).
$$
Since we have expressed the evolution operator in a symmetric form~(\ref{eq:symomega}) it simplifies considerations if we first consider a general expression of the type
\begin{multline*}
 \lv \phi+\delta \phi|\exp\lp\delta t \sym{a(\op\phi)b(\op p)+c(\op\phi)}\rp|\phi\rv=\\
\\
=\lv \phi+\delta \phi|\exp\lp\delta t\lp\frac{1}{2}(a(\op\phi)b(\op p)+c(\op\phi))\rp\rp\exp\lp\delta t\lp\frac{1}{2}(b(\op p)a(\op\phi)+c(\op\phi))\rp\rp|\phi\rv\\
+\O(\delta t^2)\equiv(*).
\end{multline*}
Inserting the identity operator (as represented in the eigenbasis of $\op p$) between the two exponentials, we can write the above as
\begin{multline*}
  (*)=\int\dm p\exp\lp-\imath\delta\phi_\alpha p_\alpha\delta t+\lp\frac{1}{2}(a(\phi+\delta \phi)+a(\phi))b( p)+\frac{1}{2}(c(\phi+\delta\phi)+c(\phi))\rp\rp\\+\O(\delta t^2).
\end{multline*}
Recalling that our stochastic differential equation~(\ref{eq:lang}) gives rise to paths for which $\delta\phi=\O(\sqrt{\delta t})$, we have
$$
  (*)=\int\dm p\exp\lp-\imath\delta\phi_\alpha p_\alpha+\delta t\lp a(\bar\phi)b(p)+c(\bar\phi)\rp\rp+\O(\delta t^2), \quad \bar\phi=\phi+\delta\phi/2.
$$
Using this, and the symmetric form of the evolution operator~(\ref{eq:symomega}), we can write
\be{eq:fram2}
  \lv \phi+\delta\phi|\exp(\delta t\, \op\Omega)|\phi\rv=\int\dm p \exp\lp-\delta S(\bar \phi,p)\rp+\O(\delta t^2),
\ee
where the dynamical-action difference $\delta S(\bar\phi,p)$ is given by
$$
  \delta S(\bar \phi,p)=\imath  p_\alpha \delta\phi_\alpha-\Omega_{\rm sym}( \bar \phi,p)\delta t.
$$
Since the action is quadratic in $p$ we could perform the integration over $p$ and then take the limit $\delta t\searrow 0$. To do this one has to take special care of terms divergent as $\delta t\searrow 0$, and this is done in~\cite{Graham77}. We do not require this form of the action in subsequent derivations, and do not pursue this matter further. The above formulation shows the advantages of using a symmetric representation of the evolution operator when considering stochastic differentials of the Stratonovich type.

When studying the effect of time reversal, we are faced with a further complication due to the possibility that the steady-state probability density is not time-reversal invariant. To account for this fact we split the generalised potential into two parts,
$$
  \Phi_\ss(\phi)=\Phi_\ss^{\rm ref}(\phi)+\Delta\Phi_\ss(\phi),
$$
where we demand the reference potential to be time-reversal invariant
$$
\Phi_\ss^{\rm ref}(\ti \phi)=\Phi_\ss^{\rm ref}(\phi).
$$
Considering the time-reversed path $\tilde\phi_{\delta t}(\cdot)=\ti \phi_{\delta t}(T-\cdot)$ we can write the path probability density for the time-reversed realisations as
\begin{multline*}
\Prob(\tilde\phi_{\delta t}(\cdot))=\left[\prod\limits_{k=0}^{T/\delta t-1}\lv \ti\phi(t_k)|\exp(\delta t\op\Omega)|\ti\phi(t_{k+1})\rv\right]\exp(-\Phi_\ss(\ti\phi(T)))\\
=\left[\prod\limits_{k=0}^{T/\delta t-1}\lv \phi(t_{k+1})|\exp(\delta t\exp(-\op\Phi_\ss^{\rm ref}){\tiop \Omega}^\dagger\exp(\op\Phi_\ss^{\rm ref}))|\phi(t_{k})\rv\right]\exp(-\Delta\Phi_\ss(\ti\phi(T))+\Delta\Phi_\ss(\phi(0))-\Phi_\ss(\phi(0))).
\end{multline*}
In the above we have used the time-reversal invariance of $\Phi_\ss^{\rm ref}$ in order to bring it inside the product. In the case of a system with a time-symmetric steady state we can choose $\Delta\Phi_\ss=0$, and the surface terms vanishes, while in the general case we can never eliminate the surface terms all together. Introducing the operator 
$$
  \op\sigma(\Phi)\Def \op\Omega-\exp(-\tiop\Phi)\tiop\Omega^\dagger\exp(\op\Phi),
$$
we can rewrite the above path probability density as
$$
\Prob(\tilde\phi_{\delta t}(\cdot))=\left[\prod\limits_{k=0}^{T/\delta t-1}\lv \phi(t_{k+1})|\exp(\delta t(\op\Omega-\op\sigma(\Phi_\ss^{\rm ref})))|\phi(t_{k})\rv\right]\exp(-\Delta\Phi_\ss(\ti\phi(T))+\Delta\Phi_\ss(\phi(0))-\Phi_\ss(\phi(0))).
$$
Through taking the symmetric representation of both $\op\Omega$ (\ref{eq:symomega}) and $\op\sigma(\Phi_\ss^{\rm ref})$,
$$
\op\sigma(\Phi_\ss^{\rm ref})=\Big[(2\imath \op p_\alpha -\partial_\alpha\op\Phi_\ss^{\rm ref})(\op k_\alpha^{\rm I}-(\partial_\beta -\partial_\beta\op\Phi_\ss^{\rm ref}  )\op\Xi_{\alpha\beta} ) -(\partial_\alpha-\partial_\alpha\op\Phi_\ss^{\rm ref})\op k_\alpha^{\rm R}\Big]_{\rm sym},
$$
 and using the same path-integral techniques as before, each term in the above product can be written as
\be{eq:deppe}
 \int \dm p \exp(-(\delta S(\bar \phi,p)+\delta t \sigma(\bar \phi,p))),
\ee
with
$$
\sigma(\bar \phi,p)= (2\imath  p_\alpha -\partial_\alpha\Phi_\ss^{\rm ref}(\bar \phi))( k_\alpha^{\rm I}(\bar \phi)-(\partial_\beta -\partial_\beta\Phi_\ss^{\rm ref}(\bar \phi) )\Xi_{\alpha\beta}(\bar \phi) ) -(\partial_\alpha-\partial_\alpha\Phi_\ss^{\rm ref}(\bar \phi)) k_\alpha^{\rm R}(\bar \phi).
$$
Through the variable change
$$
  p_\alpha \rightarrow p_\alpha-\imath\Xi_{\alpha\beta}^{-1}(\bar \phi)(k_\beta(\bar \phi)-(\partial_\gamma-\partial_\gamma\Phi_\ss^{\rm ref}(\bar \phi))\Xi_{\beta\gamma}(\bar \phi)),
$$
 which does not effect the measure $\dm p$, we can put~(\ref{eq:deppe}) in the form
\be{eq:bak}
  \exp(-\delta\sigma(\bar \phi))\int \dm p \exp(-\delta S(\bar \phi,p))
\ee
where
$$
  \delta\sigma(\phi)=(k_\alpha^{\rm I}(\bar \phi)-(\partial_\gamma-\partial_\gamma\Phi_\ss^{\rm ref}(\bar\phi))\Xi_{\alpha\gamma}(\bar\phi))\Xi^{-1}_{\alpha\beta}(\bar \phi)(\delta\phi_\beta-k^{\rm R}_\beta (\bar\phi)\delta t)-(\partial_\alpha-\partial_\alpha\Phi_\ss^{\rm ref}(\bar\phi))k_\alpha^{\rm R}(\bar\phi)\delta t.
$$
We have hereby eliminated the $p$ dependence from $\delta\sigma$, which therefore is well defined as a stochastic difference along the evolution of paths. 
Defining the (irreversible) vector 
$$
  g^{\rm ref}_\alpha(\phi)=k_\alpha^{\rm I}(\phi)-(\partial_\beta-\partial_\beta\Phi_\ss^{\rm ref}(\phi))\Xi_{\alpha\beta}(\phi)
$$
and taking the limit $\delta t\searrow 0$, we replace the symbol $\delta$ with the differential symbol $\d_{\rm s}$, and get
\be{eq:dsigma}
  \d_{\rm s}\sigma(\phi)= g^{\rm ref}_{\alpha}(\phi)\Xi_{\alpha\beta}^{-1}(\phi)(\d_{\rm s}\phi_\beta)^{\rm I}-(\partial_\alpha-\partial_\alpha\Phi_\ss^{\rm ref}(\phi))(\d_{\rm s}\phi_\alpha)^{\rm R}.
\ee
In the above we have defined the irreversible and reversible field differentials
$$
\ba{rcl}
(\d_{\rm s}\phi_\alpha)^{\rm I}\!\!\!\!&\Def&\!\!\!\!\!f^{\rm I}_\alpha(\phi)\d t+\Gamma_{\alpha i}(\phi)\d_{\rm s} w_i\\
(\d_{\rm s}\phi_\alpha)^{\rm R}\!\!\!\!&\Def&\!\!\!\!\!f_\alpha^{\rm R}(\phi)\d t, \quad \textrm{(note that $f^{\rm R}_\alpha(\phi)=k^{\rm R}_{\alpha}(\phi)$)},
\ea
$$
which transform under time reversal according to
$$
(\d_{\rm s}\phi_\alpha)^{\rm I}\rightarrow -(\d_{\rm s}\phi_\alpha)^{\rm I}, \quad (\d_{\rm s}\phi_\alpha)^{\rm R}\rightarrow (\d_{\rm s}\phi_\alpha)^{\rm R}.
$$
We further note that the first term in the above split up~(\ref{eq:dsigma}) is irreversible while the second term is reversible. For the total $\sigma$-creation along a path $\phi(\cdot)=\lim_{\delta t\searrow 0}\phi_{\delta t}(\cdot)$,
$$
  \sigma[\phi(\cdot)]\Def\int_{t=0}^{t=T} \d_{\rm s} \sigma(\phi(t)),
$$ 
it now follows from~(\ref{eq:fram2}) and~(\ref{eq:bak}) that,
\be{radon}
 \sigma[\phi(\cdot)]=\ln\left(\frac{\Prob[\phi(\cdot)]}{\Prob[\tilde\phi(\cdot)]}\right)+\Delta\Phi_\ss(\phi(0))-\Delta\Phi_\ss(\tilde\phi(0)), \quad \sigma[\tilde\phi(\cdot)]=-\sigma[\phi(\cdot)].
\ee
Thus we see that the probability density for time reflected paths are related through a simple integration of a differential along the paths. This will later directly give the Gallavotti-Cohen symmetry. We here also note that a change in reference potential $\Phi_\ss^{\rm ref}\rightarrow \Phi_\ss^{\rm ref}+\Psi$, adds the differential of the extra potential function to the $\sigma$-differential,
$$
  \d_{\rm s}\sigma(\phi)\rightarrow \d_{\rm s}\sigma(\phi)+\d_{\rm s}\Psi(\phi).
$$
It is interesting to view~(\ref{eq:dsigma}) in the light of the potential conditions~(\ref{eq:pot}). Considering the conditions for a system to exhibit detailed balance with respect to the generalized potential $\Phi_\ss^{\rm ref}$, we have
$$
  j_{\alpha}^{\rm I}(\phi)=0, \quad (\partial_{\alpha}-\partial_{\alpha}\Phi_\ss^{\rm ref})k_{\alpha}^{\rm R}(\phi)=0.
$$
If one further notes that in this case we have $g^{\rm ref}_\alpha=j_\alpha^{\rm I}$, then the direct link between the $\sigma$-creation and the breaking of the detailed-balance condition on the level of the individual paths becomes explicit. On the operator level this is also seen by the fact that the detailed balance condition~(\ref{eq:dbop}) gives $\op\sigma(\Phi_\ss^{\rm ref})=0$. We further see that the non-vanishing of the irreversible and reversible terms in the $\sigma$-creation originate in breaking the detailed-balance condition for the irreversible and reversible currents respectively.
\subsection{The Gallavotti-Cohen symmetry}
The Gallavotti-Cohen symmetry is a symmetry of the moment generating functional of $\sigma[\phi(\cdot)]$,
\be{eq:genf}
  e(\lambda)\Def-\frac{1}{T}\ln\lv \exp(-\lambda\sigma[\phi(\cdot)])\rv.
\ee
This can be put in the operator language by defining the operator $\op L_{\lambda}$ as
$$
  \lv \phi'|\op L_\lambda|\phi\rv=\lv \exp\lp-\lambda\int_{0}^{1}\d \sigma(\phi(t))\rp\rv_{\phi(0)=\phi,\,\,\phi(1)=\phi'},
$$
where the average is over paths evolving for a time $1$, given the initial and final values indicated in the subscript. Any other finite time interval could have been chosen, but we use this as it gives the simplest notation. The moment generating functional of $\sigma[\phi(\cdot)]$ can now be written as,
\be{eq:logmom}
  e(\lambda)=-\frac{1}{T}\ln\lv 0|\op L_{\lambda}^T |0\rv.
\ee
 Equivalently, since there is a unique time reversed realization corresponding to every realization, we could have taken the average over the time reversed ensemble. Using~(\ref{radon}), the moment generating functional can thus be written as
\be{eq:alt}
  e(\lambda)=-\frac{1}{T}\ln\lv 0| \exp( -\Delta\Phi_{\ss}(\tiop \phi))\op L_{1-\lambda}^T \exp(\Delta\Phi_{\ss}(\op \phi))|0\rv.
\ee
 If the steady state is time symmetric we can choose $\Delta\Phi_\ss=0$, and then equation (\ref{eq:logmom}) and (\ref{eq:alt}) directly give the Gallavotti-Cohen symmetry,
$$
  e(\lambda)=e(1-\lambda).
$$
In contrast to previously derived versions of this symmetry the above is valid for any finite time $T$.

If, on the other hand, the steady-state probability density is not time-reversal invariant, we can only show the above under certain assumptions on the dynamics and in the long time limit. To this end we assume that the dynamics is such that there is a unique, real maximal eigenvalue $\nu_{\rm{max}}(\lambda)$, corresponding to a positive eigenvector of $\op L_\lambda$ (in the finite and fully connected phase space this would be guaranteed by Perron-Frobenius theorem). First considering the average in the form~(\ref{eq:logmom}), we note that since the maximal eigenvector is positive, there is always a non-zero overlap with both $\lv 0|$ and $|0\rv$. Thus
$$
  e(\lambda)=-\nu_{\rm max}(\lambda)+\O(1/T).
$$
We could as well have considered~(\ref{eq:alt}), giving
$$
  e(\lambda)=-\nu_{\rm{max}}(1-\lambda)+\O(1/T).
$$
Combining these two results we directly get the Gallavotti-Cohen symmetry in the long time limit,
$$
  e(\lambda)=e(1-\lambda)+\O(1/T).
$$
\subsection{The fluctuation theorem}
We here illustrate the connection between the Gallavotti-Cohen symmetry and the fluctuation theorem in the case where we can choose $\Delta \Phi_\ss=0$. The situation is again more complicated for a system with a time asymmetric steady state, but is covered by the discussion in~\cite{Lebowitz99}. Denoting the average $\sigma$-creation over the time interval $[0,T]$ by $\bar \sigma$, we use~(\ref{eq:genf}) to rewrite the probability density over $\bar \sigma$ as
\begin{multline*}
  \Prob(\bar \sigma)=\int \D \phi(\cdot) \Prob[\phi(\cdot)]\delta(T\bar \sigma-\sigma[\phi(\cdot)])=\int_{-\imath \infty}^{\imath\infty}\frac{\d s}{2\pi\imath}\exp(T(s\bar \sigma-e(s)))\\
=\int_{-\imath \infty}^{\imath\infty}\frac{\d s}{2\pi\imath}\exp(T(s\bar \sigma-e(1-s)))=\exp(T\bar \sigma)\Prob(-\bar \sigma),
\end{multline*}
which is a version of the fluctuation theorem that is exact for finite $T$. 

\section{The $\sigma$-creation in two special cases}
Below we will study the form of~(\ref{eq:dsigma}) for general systems with time-reversal invariant steady-state probability densities, as well as for a Hamiltonian system coupled to a non-uniform heat reservoir and driven by non-potential forces. In the latter case we can no longer assume the steady-state probability densities to be time-reversal invariant.

In both cases we need to choose the appropriate reference potential, $\Phi^{\rm ref}_\ss$. We shall try to make the simplest choice possible, but also remember that we ultimately want to consider an analogy between~(\ref{eq:dsigma}) and the phase-space contraction rate for deterministic systems. Thus we would like the $\sigma$-differential to vanish when the steady state is approaching an equilibrium steady state. Therefor we require $\Phi_\ss^{\rm ref}=\Phi_\ss$ in equilibrium. 
\subsection{Systems with time-reversal invariant steady states}
As has been shown above, the Gallavotti-Cohen theorem can be satisfied for finite times as soon the steady-state probability density is time-reversal invariant. For a general steady state the reversible and irreversible currents must both be divergence free,
$$
  \partial_\alpha J^{\rm I}_\alpha(\phi)=\partial_\alpha J^{\rm R}_\alpha(\phi)=0.
$$
Demanding that the steady state is time-reversal invariant, and making the simplest choice $\Phi^{\rm ref}_\ss=\Phi_\ss$, we have,
$$
 \ba{rl} J_{\alpha}^{\rm R}(\phi)&=j_\alpha^{\rm R}(\phi)\rho_\ss(\phi)\\ J_{\alpha}^{\rm I}(\phi)&=j_\alpha^{\rm I}(\phi)\rho_\ss(\phi)\ea, \quad \textrm{with} \quad \ba{rl}j^{\rm R}(\phi)&=k_\alpha^{\rm R}(\phi)\\  j^{\rm I}(\phi)&=g_\alpha^{\rm ref}(\phi)=k_\alpha^{\rm I}(\phi)-(\partial_\beta-\partial_\beta\Phi_\ss(\phi))\Xi_{\alpha\beta}(\phi).\ea
$$
This results in that only the irreversible part of the $\sigma$-creation contributes
$$
  \d_{\rm s}\sigma(\phi)=j_{\alpha}^{\rm I}(\phi)\Xi_{\alpha\beta}^{-1}(\phi)(\d_{\rm s}\phi_\beta)^{\rm I}.
$$
From this it follows that 
$$
  \lv \d_{\rm s}\sigma(\phi)\rv=\lv j^{\rm I}_\alpha(\phi)\Xi_{\alpha\beta}^{-1}(\phi)j^{\rm I}_{\beta}(\phi)\rv \d t.
$$
Since we have assumed the noise-correlation matrix to be positive definite the above is always non-negative, and only vanishes in equilibrium. Thus, since the differential $\d_{\rm s}\sigma$ satisfies the fluctuation theorem, vanishes for equilibrium systems, and is on average strictly non-negative in a non-equilibrium steady state, we identify this as corresponding to the phase-space contraction rate in the deterministic (see introduction), and therefore also as the Gibbs-entropy creation rate. It further satisfies the fluctuation theorem for finite times.

In the next section we consider a important example when it is not possible to choose $\Delta\Phi_\ss=0$.
\subsection{A Langevin system with inertia}
In this section we show how the general form~(\ref{eq:dsigma}) directly recreates the results previously derived by ansatz in~\cite{Lebowitz99}. It further points to the significance played by the Hamiltonian over temperature as a reference potential. We  consider a Hamiltonian system coupled to a heat bath of inverse temperature $\beta({\bm \pos})$, and with the Hamiltonian
$$
  H({\bm \pos},{\bm \kin})=\frac{1}{2}{\bm \kin}\cdot{\bm \kin}+V({\bm \pos}).
$$
We further assume the system to be influenced by a non-potential driving force ${\bm f}^{\rm drive}({\bm \pos})$, and a viscous drag $-\gamma(\bm \pos)\bm \kin$. The fields are taken to evolve according to the Langevin equation
$$
\begin{pmatrix}\d_{\rm s}{\bm \pos}\\\d_{\rm s}{\bm \kin}\end{pmatrix}=\begin{pmatrix}\frac{\partial H({\bm \pos},{\bm \kin})}{\partial {\bm \kin}}\d t+\epsilon\d_{\rm s}{\bm w}_{\pos}\\ \lp -\frac{\partial H({\bm \pos},{\bm \kin})}{\partial {\bm \pos}}+{\bm f}^{\rm drive}({\bm \pos})-\gamma({\bm \pos}){\bm \kin}\rp \d t +\sqrt{2\gamma(\bm\pos)/\beta(\bm\pos)}\d_{\rm s}{\bm w}_{\kin} \end{pmatrix},
$$ 
where we have introduced an small noise term in the $\bm \pos$-differential in order to keep the noise-correlation matrix non-singular. We will keep $\epsilon$ finite for now, but set it to zero at the end. The specific form of the stochastic force acting on the generalized-momentum coordinates is chosen to given the equilibrium steady state
$$
  \rho_{\rm eq}({\bm \pos},{\bm \kin})\propto \exp(-\beta H({\bm \pos},{\bm \kin})),
$$
in the absence of any drive and temperature gradient. The reversible and irreversible drifts become
$$
 ( k_\alpha^{\rm R})=\begin{pmatrix}{\bm \kin} \\ -\frac{\partial  V({\bm q})}{\partial {\bm \pos}}+{\bm f}^{\rm drive}({\bm \pos})\end{pmatrix}, \quad ( k_\alpha^{\rm I})=\begin{pmatrix}0\\ -\gamma({\bm q}){\bm\kin}\end{pmatrix}.
$$
For general position dependent $\beta(\bm \pos)$ and with an (non-potential) drive present, the steady-state potential is in general not time-reversal invariant. We then have to chose some other reference potential. The most natural choice seems to be the equilibrium potential, and this is indeed the choice that recreates the results of~\cite{Kurchan98,Lebowitz99}.
So, using 
$$
 \Phi_\ss^{\rm ref}(\bm\pos,\bm\kin)=\beta(\bm \pos)\H(\bm\pos,\bm\kin) \quad \textrm{and}\quad \bm \Xi(\bm \pos,\bm \kin)=\begin{pmatrix}\epsilon^2/2&0\\ 0& \gamma(\bm\pos)/\beta(\bm \pos)\end{pmatrix},
$$
we have
\be{eq:hamdsig}
  \d_{\rm s}\sigma(\bm\phi)=\O(\epsilon)+\beta(\bm \pos)\lp H({\bm \pos},{\bm \kin})\frac{\partial  \ln \beta({\bm \pos})}{\partial {\bm \pos}}+ {\bm f}^{{\rm drive}}({\bm \pos})\rp \cdot {\bm \kin}\d t.
\ee
It is now safe to take the limit $\epsilon\searrow 0$, and contrary to the situation for the systems with a time-reversal invariant steady-state probability density, all the $\sigma$-creation is reversible.

 Since the steady state is not time symmetric, the Gallavotti-Cohen symmetry and the fluctuation theorem are asymptotically valid only in the long-time limit. It further follows that the $\sigma$-creation integrated along the paths and averaged over histories is asymptotically non-negative as one considers longer and longer times~\cite{Lebowitz99}. Equation~(\ref{eq:hamdsig}) is the same result as in \cite{Kurchan98,Lebowitz99}, while it is re-derived here to illustrate that it is contained in the general form~(\ref{eq:dsigma}). It further underlines the difference in choice of reference potential compared with the case where the system displays a time-symmetric steady state.

In~\cite{Lebowitz99} the above differential~(\ref{eq:hamdsig}) was interpreted as work done on the system by the driving force and the temperature gradient, divided by temperature. 
\section{Conclusion}
For general Langevin systems we have derived a general form~(\ref{eq:dsigma}) of the stochastic differential that satisfies the Gallavotti-Cohen symmetry. 
This was done by considering the effect of time reversal on the path probability densities generated by the dynamics.

In the general form~(\ref{eq:dsigma}) we have introduced a time-reversal symmetric reference potential $\Phi_{\rm ref}$. In the case when the dynamics gives rise to a time-symmetric steady state, we can choose $\Phi_{\rm ref}=\Phi_\ss$, and the $\sigma$-creation satisfies the fluctuation theorem for finite times. It also vanished along every path for equilibrium systems, while being on average strictly positive for non-equilibrium systems. By analogy with deterministic systems we thus identify $\int\d_{\rm s}\sigma(\phi(t))$ as the Gibbs-entropy creation along the path $\phi(t)$. 

We also considered the special case of a Hamiltonian system with an non-uniform temperature field, and driven by a non-potential force. Here the steady-state probability density can not be assumed time-reversal invariant, and a different reference potential has to be chosen. By considering the Hamiltonian over temperature as the reference potential, we were able to recreate previous results~\cite{Kurchan98,Lebowitz99}, giving a differential that also can be interpreted as some form of entropy production. 

\section{Acknowledgements}
I would like to thank Robin Stinchcombe for several useful suggestions, and  Joel L. Lebowitz and Herbert Spohn for crucial comments on an early versions of the manuscript.

\bibliography{bibliography}
\end{document}